\documentclass[twocolumn,preprintnumbers,amsmath,amssymb]{revtex4}
\usepackage{graphicx}
\usepackage{dcolumn}
\usepackage{bm}

\newcommand{\comment}[1]{}

\begin{document}

\title{Probing the quantum state of a guided atom laser pulse}
\author{Kevin L. Moore}
\email{klmoore@berkeley.edu}
\author{Subhadeep Gupta}
\author{Kater W. Murch}
\author{Dan M. Stamper-Kurn}

\affiliation{Department of Physics, University of California,
Berkeley CA 94720}

\date{\today }

\begin{abstract}
We describe bichromatic superradiant pump-probe spectroscopy as a
tomographic probe of the Wigner function of a dispersing particle
beam.  We employed this technique to characterize the quantum state
of an ultracold atomic beam, derived from a $^{87}$Rb Bose-Einstein
condensate, as it propagated in a $2.5 \,$mm diameter circular
waveguide. Our measurements place an upper bound on the longitudinal
phase-space area occupied by the $3 \times 10^5$ atom beam of $9(1)
\hbar$ and a lower bound on the coherence length (${\cal L} \geq
13(1) \, \mu$m). These results are consistent with full quantum
degeneracy after multiple orbits around the waveguide.

\end{abstract}


\maketitle


Advances in the control of quantum degenerate gases have mirrored
those of optical lasers, including the realization of high-contrast
atom interferometers \cite{gupt02inter,torii00mach}, nonlinear atom
optics \cite{deng99} and dispersion management \cite{eier03,eier04}.
Further, given single-mode waveguides for atoms \cite{lean02guide}
and other atom optical elements, the prospect of sensitive
guided-atom interferometry has invited intensive experimental
pursuit.  Critical to realizing this prospect are methods for
characterizing the coherence of a guided atom beam, analogous to
beam characterization in a high-energy particle accelerator.


Pulsed particle beams are naturally described by the Wigner
quasi-probability distribution, defined as \cite{wign32}
\begin{equation}
{\cal W} \left ( {\bf r}, {\bf p} \right ) = \frac{1}{2 \pi} \int \,
e^{-i {\bf p} \cdot {\bf y} / \hbar} \langle {\bf r} - \frac{{\bf
y}}{2} | \hat{\rho} | {\bf r} + \frac{{\bf y}}{2} \rangle \, d {\bf
y},
\end{equation}
with $\hat{\rho}$ being the density matrix of the system.  This
distribution is the quantum mechanical equivalent of the classical
phase-space distribution.  Experimentally, ${\cal W} \left ( {\bf
r}, {\bf p} \right )$ is determined tomographically by measuring its
projection at various angles in phase space
\cite{leib96motion,kurt97}.


In this Letter we describe the use of bichromatic superradiant
pump-probe spectroscopy (SPPS) to quantify the coherence of a
dispersing atomic beam propagating in a circular waveguide
\cite{gupt05tort}. We show how the waveguide curvature allows for
tomographic measurements of the Wigner function of the beam and,
thereby, for an accurate measurement of its phase-space density.
Both long-range coherence and single transverse mode propagation
were evident over many revolutions of atoms in the waveguide,
implying that a guided atom laser pulse derived from a Bose-Einstein
condensate remains coherent for at least $300 \,$ms of propagation.

Superradiant light scattering from quantum degenerate gases provides
striking confirmation of their long-range coherent nature
\cite{inou99super2,yosh05super}. An elongated cloud undergoing
superradiance scatters light preferentially into ``end-fire modes,''
leading to highly directional emission \cite{rehl71}. Coherence
between scattered and unscattered atoms establishes a periodic
grating of density or polarization which stimulates further light
scattering. Once established, whether by superradiance or otherwise
\cite{saub97,inou00amp}, this grating will decay or dephase on a
timescale $\tau_c = m /2 |\bf{q}| \sigma_p$ with $m$ being the
atomic mass, $\hbar \bf{q}$ the superradiant scattering recoil
momentum and $\sigma_p$ the rms momentum spread of the unscattered
atoms along the recoil direction. The decay of coherence can be
isolated experimentally by applying superradiance in a pump-probe
manner. After a first optical pump pulse initiates superradiance and
establishes coherence in the gas, this coherence is allowed to decay
freely for a time $\tau$ before a second optical pulse is applied.
In Ref.\ \cite{yosh05super}, this pump-probe technique revealed in
detail the bimodal momentum distribution of a partly-condensed Bose
gas.

\begin{figure}
       \includegraphics[width=3.5in,clip]{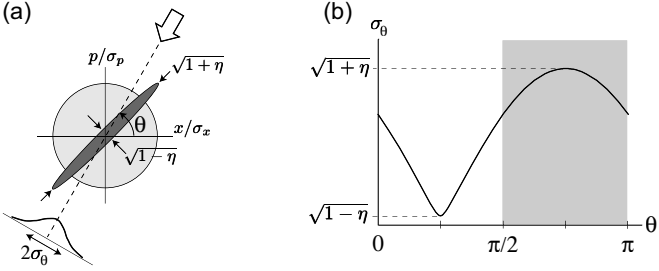}
        \caption{Projective measurements as probes of quantum
        degeneracy.
        (a) Contours of Gaussian Wigner distributions ${\cal W}(x,p)$ are shown.
        ${\cal W}(x,p)$ is determined
        by its projections at all angles
        $0 \leq \theta < \pi$.  Measurements of only the momentum and
        position distributions ($\theta = 0$ and $\theta = \pi/2$ projections,
        respectively),  cannot distinguish a homogeneous (light shading)
        from a correlated ensemble (dark shading).  (b)
        rms widths of distributions derived at various projection
        angles are shown.  Time-of-flight analyses recover a limited range of projection angles
        (shaded),
        while bichromatic SPPS accesses all projection
        angles.
        \label{skewedprobe}}
    \end{figure}

Let us consider an elongated beam of $N$ atoms in the transverse
ground state of a 1D waveguide with longitudinal rms spatial and
momentum widths of $\sigma_x$ and $\sigma_p$, respectively. The 1D
Wigner function of the beam is bounded by these widths to occupy a
phase-space area no larger than $\mathcal{A}_{max} = \sigma_x
\sigma_p $. However, $\mathcal{A}_{max}$ may represent a gross
overestimate of the actual phase space area occupied by the beam.
For example, consider that the beam originates from a
thermally-equilibrated trapped gas that was released into the
waveguide.  Free expansion of the gas causes the momentum and
position of the beam to be strongly correlated, a feature captured
by a posited Wigner function of the form
\begin{equation}
{\cal W}\left( x,p\right) = \frac{\exp\left[ - \frac{1}{2 (1 -
\eta^2)} \left( \frac{x^2}{\sigma_x^2} - 2 \eta \frac{ x p
}{\sigma_x \sigma_p} + \frac{p^2}{\sigma_p^2} \right) \right] }{\pi
\sigma_x \sigma_p \sqrt{1 - \eta^2}}\label{eq:wignerposit}
\end{equation}
where $\eta = \langle p \, x \rangle / \sigma_p \sigma_x$  (Fig.
\ref{skewedprobe}). The actual phase-space area $\mathcal{A}$
occupied by such a beam is smaller than the aforementioned estimate
by a factor $\sqrt{1 - \eta^2}$. That is, for proper
characterization of an atomic beam one must distinguish between a
spatially-inhomogeneous momentum width $\sigma_p$, which may be
dominated by a coherent velocity chirp across the length of the
beam, and a ``homogeneous'' width $ \mathcal{A} /\sigma_x$.

\begin{figure}
        \includegraphics[width=3.5in,clip]{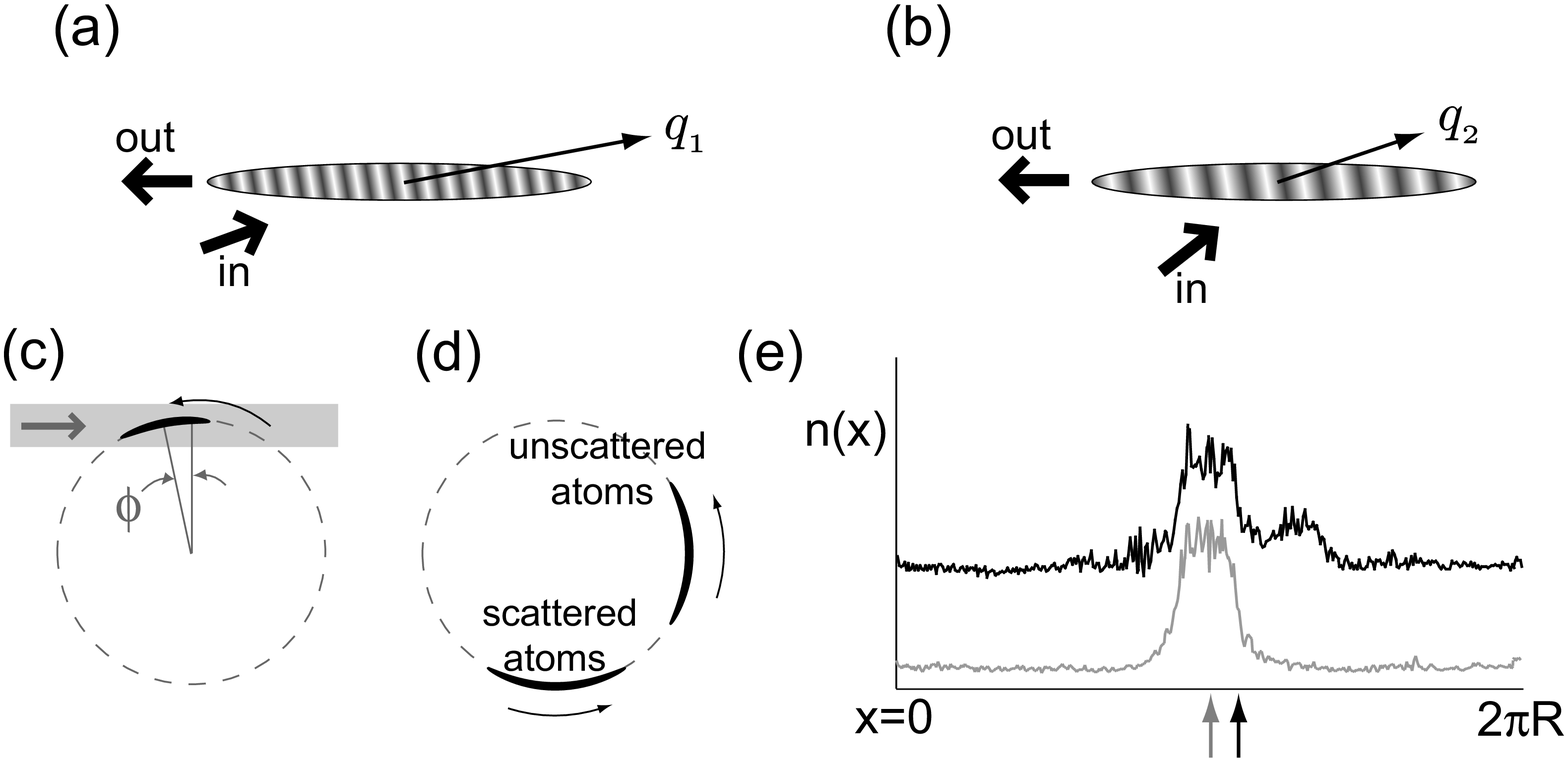}
        \caption{Bichromatic SPPS
        in a circular waveguide.
        (a) Superradiant Rayleigh scattering of a pump pulse establishes
        a density modulation of wavevector $\hbar \bf{q}_1$ in an elongated atomic beam.
        (b) A coherent velocity chirp causes the modulation wavevector to decrease along
        the long axis.  The remaining coherence is revealed
        by light scattering with recoil momentum $\hbar \bf{q}_2$ matched to the
        modified density grating.
        (c) Pump (probe) light illuminates the freely propagating atom beam
        at angle $\phi$ ($\phi + \Omega \tau$) relative to the mean angular
        position, and (d) scattered atoms separate
        from the original pulse and can be distinguished from
        unscattered atoms.
        (e) Azimuthal density distributions
        $n(x)$ in the ring 160 ms after illumination are shown for beams that have (black) or have not
        (grey) undergone superradiant light scattering.  The shifted center of mass
        (indicated by arrows) quantifies the total superradiant scattering rate.
        \label{fig2}}
    \end{figure}

To access these correlations, we consider \emph{bichromatic} SPPS in
which the recoil momenta $\hbar \textbf{q}_1$ and $\hbar
\textbf{q}_2$ imparted by superradiance are different for the pump
and probe pulses, respectively (Fig.\ \ref{fig2}a-b). These
differing momenta may result experimentally from pump and probe
pulses which differ in wavevector, or, as in the present experiment,
which differ in their angle of incidence with respect to the long
axis of the cloud. Restricting our treatment to one dimension along
$\hat{x}$, the superradiant scattering rate $\Gamma$ from the second
(probe) light pulse \cite{cour96excitations,inou99super2} can be
expressed in terms of the Wigner function of the state of the system
\emph{before} the first (pump) pulse as
\begin{equation}
\Gamma \propto \left| \int \!\!\! \int e^{ i \left( \frac{q_1
\tau}{m} p + \Delta q  x \right)} \, {\cal W}(x,p) \, d x\, d p
\right|^2 . \label{phasematch}
\end{equation}
where $\Delta q = q_2 - q_1$ with $\hbar q_1$ and $\hbar q_2$
being the projections of the recoil momenta along the $\hat{x}$
axis, and $\tau$ is the pump-probe delay time. Performing an
extended canonical transformation to generalized coordinates
$\tilde{x} = (x/\sigma_x) \cos\theta + (p/\sigma_p) \sin\theta$
and $\tilde{p} = - (x/\sigma_x) \sin\theta + (p /\sigma_p)
\cos\theta$, with $\tan\theta = -\frac{\Delta q m}{q_1 \tau}
\frac{\sigma_x}{\sigma_p}$ we obtain
\begin{equation}
\Gamma \propto \left|  \int  \, \,  e^{ i \left( \frac{\sigma_p q_1
\tau}{m} \cos \theta  - \sigma_x \Delta q \sin \theta \right)
\tilde{p}} \, d \tilde{p} \, \int \, {\cal W}(x,p) \, d \tilde{x}
\right|^2 . \label{otherphasematch}
\end{equation}
Monochromatic SPPS ($\Delta q = 0$) yields information only on the
overall momentum distribution of the atomic system, which derives
from projecting the Wigner function on the momentum axis ($\theta =
0$) \cite{yosh05super}.  In contrast, bichromatic SPPS assesses the
Wigner function at a non-zero projection angle $\theta$. In
particular, tuning experimental parameters such that $\theta =
\pi/4$ probes the Wigner function of Eq.\ \ref{eq:wignerposit} along
the narrow axis corresponding to the linear momentum chirp across
the cloud, and thereby provides a sensitive measurement of $\eta$
and of the phase-space density of the beam.

In other words, in monochromatic SPPS the reduction of the
superradiant scattering rate from a linearly-chirped beam comes
about mainly by dephasing. The density modulation established by the
pump pulse evolves at a frequency which is Doppler-shifted upward on
one end and downward on the other end of the momentum-chirped beam.
Over time, this causes the wavevector of the density modulation to
decrease linearly with time.  In bichromatic SPPS, by matching the
recoil momentum of the probe pulse to the wavevector of the density
grating, we recover a superradiant scattering rate which reveals the
remaining homogeneous decay of motional coherence.

We now turn to our implementation of this scheme to probe a pulsed
atom laser beam in a circular waveguide.  This beam originated from
a $^{87}$Rb Bose-Einstein condensate of $3 \times 10^5$ atoms
produced in a magnetic time-orbiting ring trap (TORT)
\cite{gupt05tort,murc06betatron}, biased to yield a
three-dimensional harmonic trap with trapping frequencies
$(\omega_{x}, \omega_T) = 2 \pi \times (35, 85) \, \mbox{s}^{-1}$ in
the axial (i.e.\ azimuthal in the ring) and transverse directions,
respectively. These atoms were launched azimuthally by adiabatically
decompressing the trap to $\omega_{x} = 2 \pi \times 6 \,
\mbox{s}^{-1}$ and displacing the trap minimum to a new longitudinal
position for 30 ms, accelerating the cloud to a mean orbital angular
frequency $\Omega = 2 \pi \times 8.4 \, \mbox{s}^{-1}$ chosen to be
far from any betatron resonances \cite{murc06betatron}. The TORT
potential was then balanced over the next 30 ms and operated with
radius $R = 1.25 \,$mm and $\omega_T$ as above.  The launched atomic
beam was allowed to propagate freely in this circular guide.

While in the particular case examined, the beam's provenance as a
Bose-Einstein condensate suggests its full coherence at later times,
it may also be argued that heating from trap vibrations and
imperfections, collisions with background gas particles, or effects
related to the quasi-one-dimensional nature of the waveguided atoms
\cite{bong01guide} can indeed cause the coherence to be spoiled
after sufficient propagation times. Thus, our experimental goal was
to  measure quantitatively the coherence of this propagating atom
beam at an arbitrary time after its launch.

We made use of direct absorption imaging of the propagating atom
beam to discern several properties of its evolution.  Such imaging,
applied along the symmetry axis of the circular waveguide,
quantified the longitudinal linear density of the beam $n(x)$ (Fig.
\ref{fig2}c). From the growth of the spatial width $\sigma_x$ of the
beam vs.\ propagation time, we determined the rms momemtum width as
$\sigma_p = m \times 1.8$ mm/s, a value within 10\% of that expected
due to the release of interaction energy in the launched
Bose-Einstein condensate.  This rms momentum width is obtained after
$\approx 10$ ms of mean-field acceleration, and would be expected to
generate an atom beam highly correlated in momentum and position
($\eta \simeq 1$). The transverse state of the atomic beam was
characterized by suddenly releasing the atom beam from the waveguide
and imaging the transverse extent of the beam after variable times
of flight. These observations agreed well with a mean-field model of
the coherent expansion of a Bose-Einstein condensate into a tight
waveguide \cite{sala02}, and indicated the transverse state of the
beam to be the ground state of the harmonic transverse confining
potential after about 100 ms of propagation. The beam can thus be
treated as one dimensional with its azimuthal state remaining
unknown. Combining these observations, we obtain an upper bound on
the longitudinal phase-space area of $\mathcal{A}_{max} = 310 \hbar$
for the beam after a half-revolution in the guide given its
$\sigma_x = 120 \, \mu$m rms width at that stage.

This constraint on the phase-space area was greatly improved by
application of SPPS to the propagating atom beam.  For this, the
probe and pump pulses were \emph{both} obtained from a single beam
propagating in the plane of the waveguide (to within $\pm 1^\circ$),
with a $0.4$ mm beam diameter, a detuning $560\,$MHz below the $^2
S_{1/2}$, $F=1$ $\rightarrow$ $^2 P_{3/2}$, $F=0$ transition, and
circular polarization. A typical intensity of $10$ mW/cm$^2$ was
used, corresponding to observed single-particle Rayleigh scattering
rates of $400 \, \mbox{s}^{-1}$, and pulses were typically $50 \,
\mu$s in duration. After application of the superradiant pulses, the
atoms were allowed to propagate further in the waveguide until the
scattered atoms had clearly separated from the unscattered atoms
(Fig.\ \ref{fig2}c-e). The fraction of scattered atoms and, hence,
the total superradiant scattering rate from the pump-probe sequence,
was then determined from the center of mass of the beam $x_{CM}$ in
the azimuthal coordinate .  This approach has the benefit of being
unaffected by atom-atom collisions in the guided cloud.

    \begin{figure}[htb]
        \includegraphics[width=0.5\textwidth]{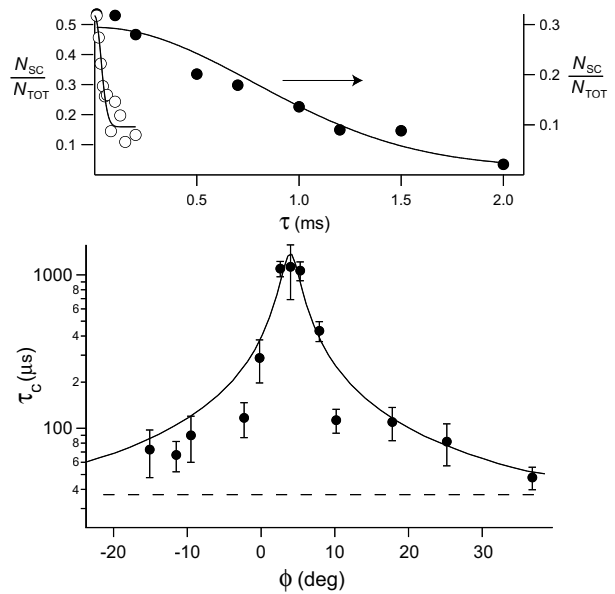}
        \caption{Bichromatic SPPS of a quantum degenerate
        beam at approximately a half revolution in the circular waveguide.
        (a) SPPS at $\phi = 38^{\circ}$ (open circles) and $\phi = 4^\circ$
        (closed circles) gives coherence times
        $\tau_c = 47 (8) \, \mu$s and $1.1(1) \,$ms, respectively,
        defined by the $1/e$ decay time of the
        superradiant signal (Gaussian fits to data are shown).
        (b) Measured coherence times are compared to theoretical
        predictions for a coherent Gaussian beam
        (solid line) and an incoherent, uncorrelated ensemble (dotted
        line).  The theoretical curve in fact predicts the maximum
        coherence time at $\phi = 31^\circ$ (see text), but has
        been shifted for comparison to data.
        \label{latestangle}}
    \end{figure}

Such pump-probe spectroscopy was applied to the atom beam at
different propagation times, and thus at different locations in the
circular guide.  As shown in Fig.\ \ref{latestangle}, the measured
coherence times depend strongly the position of the beam in the
guide. Letting $\phi$ measure the central angular position of the
beam away from the point at which the pump/probe light is tangential
to the guide, the superradiant response of atoms at large angles
($|\phi | \gtrsim 20^\circ$) decays after a pump-probe delay time of
around $\tau = 50 \, \mu$s, consistent with the coherence time
discussed above for monochromatic SPPS determined by the overall
momentum width of the beam.  In contrast, for beam positions closer
to $\phi = 4^\circ$, the coherence time is dramatically increased to
over $1 \,$ms, indicating coherence in the beam beyond that implied
solely by the overall momentum width. Similar coherence times were
observed after one, two, and three full revolutions around the ring.

This strong geometric dependence can be understood in the context of
bichromatic SPPS.  During the time $\tau$ between application of the
pump and probe pulses, the propagating atom beam rotates by an angle
$\Omega \tau$, thereby varying the relative orientation between the
incident light and the end-fire superradiant emission from the gas.
Thus, expressed in a frame co-rotating with the atom beam, the
superradiant recoil momenta of the pump and probe beams differ by
$\Delta \textbf{q} \simeq k \Omega \tau \left( - \sin \phi \hat{x} +
\cos \phi \hat{r} \right)$, with $\hat{x}$ and $\hat{r}$ being unit
vectors in the azimuthal and radial transverse directions,
respectively, and assuming $\Omega \tau \ll 1$.

With these approximations we apply the one-dimensional treatment of
bichromatic SPPS to this situation by considering just the
contribution of longitudinal phase matching to superradiant
scattering.  SPPS applied to the rotating beam while at an angle
$\phi$ probes the Wigner function of the beam at a \emph{constant}
phase-space projection angle given by $ \tan \theta = \frac{m \Omega
\sigma_x}{\sigma_p} \frac{\sin \phi}{1 + \cos\phi}$.

Using experimentally-measured quantities for the beam after a half
revolution in the waveguide, the condition $\tan \theta = 1$ for
probing the homogeneous momentum width of the correlated atom beam
is predicted to occur at $\phi_c = 31^\circ$. This value clearly
does not match the experimentally observed $\phi_c = 4(2)^\circ$
(Fig.\ \ref{latestangle}b).  2D models which numerically evaluated
the superradiance phase matching integral \cite{inou99super2} showed
that the beam curvature alone did not resolve this disagreement.
Rather, to account for this discrepancy, we suspect it is necessary
to adapt our 1D treatment of superradiance to beams with small
Fresnel number, i.e. with length greatly exceeding the Rayleigh
range defined by the probe wavelength and the transverse width of
the atom beam.  We suspect that our method may be probing only short
portions of the beam, the momentum width of which is enhanced by
their small extent, rather than probing the beam as a whole.

\comment{ , implying the assumptions made for the evolution of a
coherent beam in a curved circular waveguide system do not
immediately correspond to the 1D theory used to derive Eq.\
\ref{otherphasematch}. Given that length of the beam exceeds the
Rayleigh range \cite{rayleighfootnote} by more than an order of
magnitude, the postulate that the initial light pulse induces an
outgoing plane wave is particularly suspect. Resolving this
discrepancy will be the subject of further theoretical
investigations.}

Despite the imperfect match between the 1D theory and the
experimental data, the most important prediction of bichromatic SPPS
in a rotating system --- long coherence times at $\phi < 0$ --- is
clearly evident in this system.  We thus assert that the
observations retain their relevancy as a probe of the phase-space
distribution of the atom beam.  From the maximum coherence time of
$\tau_c = 1.1(1) \,$ms, we obtain an empirical value of $\eta = 1 -
(4.9(6) \times 10^{-4})$ for the aforementioned correlation
parameter. The atom beam is thus constrained to inhabit a
phase-space area of no more than ${\cal A} = 9(1) \hbar$ ,
equivalent to placing a lower bound of ${\cal L} = \left( \hbar
|{\bf q}|/m \right) \tau_c = 13(1) \, \mu$m \cite{saub97} on the
longitudinal coherence length of the propagating cloud.

The maximum coherence time observed is plausibly limited not by the
lack of longitudinal coherence, but rather by the decay of the
superradiant scattering rate $\Gamma(\tau)$ due to transverse phase
matching. Assessing a 2D phase-matching integral with the transverse
state being the non-interacting ground state of the transverse
trapping potential, one finds an upper bound on the coherence time
of $\approx (2 \Omega k \sigma_T \cos \phi)^{-1} < 1200 \, \mu$s
with $\sigma_T = \sqrt{\hbar / 2 m \omega_T}$ and $\omega_T$ being
the transverse trap frequency. Thus, our observations should be
construed as placing quantitative lower bounds on the coherence of
the propagating atom beam while remaining consistent with its
complete coherence.

In conclusion, we have described bichromatic SPPS as a technique to
perform tomography of the Wigner function of a propagating atom
beam. We implemented this technique to probe an ultracold atomic
beam propagating in a circular waveguide and observed long coherence
times consistent with a highly degenerate quantum ensemble of $3
\times 10^5$ atoms occupying no more than $10$
($\sigma_x/\mathcal{L}$) phase-space cells.

This work was supported by DARPA (Contract F30602-01-2-0524), ARO,
and the David and Lucile Packard Foundation. KLM acknowledges
support from  NSF and SG from the Miller Institute.


\end{document}